\documentclass[12pt,aps]{revtex4}
\usepackage{amsmath,amssymb,graphicx}
\begin{document}

\title{Area-scaling of quantum fluctuations}
\author{Amos Yarom, Ram Brustein}
\affiliation{ Department of Physics, Ben-Gurion University, Beer-Sheva 84105, Israel \\
    {\rm E-mail:}
    {\tt yarom@bgumail.bgu.ac.il, ramyb@bgumail.bgu.ac.il, } }


\begin{abstract}
We show that fluctuations of bulk operators that are restricted to
some region of space scale as the surface area of the region,
independently of its geometry. Specifically, we consider two point
functions of operators that are integrals over local operator
densities whose two point functions falls off rapidly at large
distances, and does not diverge too strongly at short distances.
We show that the two point function of such bulk operators is
proportional to the area of the common boundary of the two spatial
regions. Consequences of this, relevant to the holographic
principle and to area-scaling of Unruh radiation are briefly
discussed.
\end{abstract}

\maketitle
\section{Introduction}
The discovery of the non extensive nature of black hole entropy
\cite{Bekenstein,Hawking} has lead to entropy bounds on matter,
and to the proposal of the holographic principle (see
\cite{Bousso} for a review) -- a conjecture regarding the
reduction of the number of degrees of freedom needed to describe a
theory of quantum gravity. Later, it was discovered that other
thermodynamic quantities of fields in a black-hole background also
scale as the surface area of its horizon, and consequently, it was
hypothesized that it is the entangled nature of the quantum state
of the system inside and outside the horizon, which leads to
area-scaling. This view has received some support from the
area-scaling properties of entanglement entropy in flat space
\cite{Srednicki,Bombelli,Casini}.

In this paper we study area-scaling of two point functions of a
certain class of quantum field theory bulk operators in Minkowski
space, to be defined shortly. We shall show, under weak
assumptions, that quantum expectation values of operators
restricted to a region of space will scale as its surface area,
regardless of the region's geometry. This area-scaling property
may then be used to establish area-scaling of thermodynamic
quantities \cite{TandA}, and in some sense, a bulk-boundary
correspondence \cite{HDR}.

The two point functions we wish to consider are the expectation
values of a product of two operators which are restricted to some
regions $V_1$ and $V_2$ of Minkowski space. In order to restrict
operators to a region $V_i$, we use operator densities
$\mathcal{O}_j(\vec{x})$, and define $O^{V_i}_j = \int_{V_i}
\mathcal{O}_j(\vec{x}) d^dx$.

We show that if the connected two point function of the operator
densities $F(|\vec{x}-\vec{y}|)$,
$\langle \mathcal{O}_i(\vec{x}) \mathcal{O}_j(\vec{y}) \rangle_C
    = F(|\vec{x}-\vec{y}|) \equiv \nabla^2 g
$ satisfies the following conditions:
\begin{itemize}
\label{L:one}
        \item[(I)] $g(\xi)$ is short range: at large distances it behaves like $g(\xi)\sim1/\xi^{a}$, with
        $a \geq d-1$,
\label{L:two}
    \item[(II)] $g(\xi)$ is not too singular at short distances. Explicitly,
          we require that for small $\xi$, $g(\xi) \sim 1/\xi^{a}$
          with $a<d-2$,
\end{itemize}
and if $V_1$ and $V_2$ are finite, then the connected two point
function $ \langle O^{V_1}_i O^{V_2}_j \rangle_C $ is proportional
to the area of the common boundary of the two regions $V_1$ and
$V_2$:
    $\langle O^{V_1}_i O^{V_2}_j \rangle_C
     \propto S(B(V_1) \cap B(V_2))$.
Here $B(V)$ is the boundary of $V$, and $S(B(V))$ is its area.

This implies that the fluctuations (or variance:
$\text{var}(O_i^V)=\langle({O_i^V}^2-\langle O_i^V \rangle^2
\rangle$) of the operator $O^V_i$ scale as the surface area. In
particular, the energy fluctuations inside $V$ will be
proportional to $S(B(V))$. This is discussed at length in
\cite{TandA}.

In section \ref{S:Cal_of_corr} we give a detailed proof of the
area-scaling property of two point functions of bulk operators
satisfying conditions (I) and (II). Section \ref{S:Explicit_Cal}
contains an explicit calculation of energy fluctuations, and
fluctuations of the boost operator. We discuss these results in
section \ref{S:Discussion}.

\section{Area-scaling of two point functions}
\label{S:Cal_of_corr}
We shall first give a general explanation as to why conditions (I)
and (II) are required for area-scaling, and outline the proof for
area-scaling of two point functions. A more detailed discussion
will then follow.

In order to evaluate $\langle O^{V_1}_i O^{V_2}_j\rangle_C$, we
may express it as follows,
\begin{align}
\notag
    \langle O^{V_1}_i O^{V_2}_j\rangle_C
        &=\int_V\int_V F_{i,j}(|\vec{x}-\vec{y}|) d^dx d^dy\\
\label{E:OV1OV2}
        &=\int D(\xi) F_{i,j}(\xi) d\xi.
\end{align}
The integral has been factored into a product of a purely
geometric term $D(\xi)$, and an operator dependent term
$F_{i,j}(\xi)$. Using $\nabla^2 g_{i,j}(\xi) \equiv F_{i,j}(\xi)$,
we may integrate eq.(\ref{E:OV1OV2}) by parts. The surface term
then vanishes due to conditions (I) and (II), and we get
\begin{equation}
\label{E:partshandwaving}
    \langle O^{V_1}_i O^{V_2}_j\rangle_C =
            - \int \frac{d}{d \xi} \left( D(\xi) \frac{1}{\xi^{d-1}}\right)
                   \xi^{d-1} \frac{d}{d \xi} g(\xi)
              d\xi.
\end{equation}

To proceed we note that the geometric term is of the form:
\begin{equation}
\label{E:defofDV1V2}
    D_{V_1,V_2}(\xi)=
        \int_{V_1} \int_{V_2}
        \delta(\xi-|\vec{r}_1-\vec{r}_2|)
        d^dr_1 d^dr_2.
\end{equation}
We show in the next subsection that $D_{V_1,V_2} = G_V V \xi^{d-1}
+ G_S S \xi^d +\mathcal{O}(\xi^{d+1})$, with $V$ a volume term,
$S$ a surface area term, and G numerical coefficients. There are
some geometries for which this scaling is more obvious: for
example, if $V_1$ and $V_2$ are disjoint volumes with a common
boundary which may be approximated as flat, then one may carry out
the integral in eq.(\ref{E:defofDV1V2}) by switching to `center of
mass coordinates'
    $\vec{R}=\frac{\vec{r}_1+\vec{r}_2}{2}$,
and
    $\vec{r}=\vec{r}_2-\vec{r}_1$.
Restricting the value of $|\vec{r}|$ to be equal to $\xi$, results
in confining $\vec{R}$ to be, at most, a distance of $\xi/2$ from
the boundary. The integral over the $\vec{R}$ coordinate will give
a term proportional to $S \xi$, S being the area of the boundary.
The remaining integral over  the $\vec{r}$ coordinate will give a
term proportional to $\xi^{d-1}$. Combining these results, we get
the claimed form of $D_{V_1,V_2}$ (the volume term vanishes in
this case).

\begin{figure}[btp]
  \begin{center}
  \scalebox{0.5}{\includegraphics{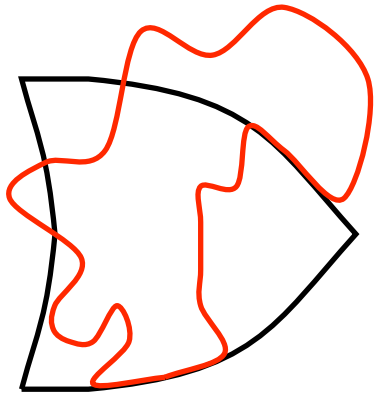}}
  \caption{A general case when the regions have some overlap with each other,
  and neither is fully contained in the other. The regions have
  common `boundaries from within', and `boundaries from the outside'.}
  \label{F:general case}
  \end{center}
\end{figure}

Another possibility is that $V_1=V_2$. In this case an integral
over the center of mass coordinate $\vec{R}$ will yield, to
leading order, a volume dependent term. The remaining integral
over the $\vec{r}$ coordinate will give the appropriate powers of
$\xi$.

Using these two simple cases, one can now construct $D_{V_1,V_2}$
for general geometries by dividing one of the volumes, say $V_1$,
into sub-volumes which are either contained in $V_2$, just
touching $V_2$, or disconnected from $V_2$. In order for the
appropriate form of $D_{V_1,V_2}$ to follow, one would need that
the coefficient $G_S$ be exactly the same for both cases described
above. This requires a more detailed calculation and is shown in
the following subsection.

Going back to eq.(\ref{E:partshandwaving}), we observe that the
integral comes from the region $\xi \to 0$, and that the
contribution of the volume dependent term of $D_{V_1,V_2}$ will
vanish $\frac{\partial}{\partial \xi}
\left(\frac{D(\xi)}{\xi^{d-1}}\right) \sim S +\mathcal{O}(\xi)$.
The leading contribution to the integral will come from the
surface term, which gives area-scaling behavior of two point
functions, as claimed. The vanishing of the contribution of the
volume term is a result of properties (I) and (II) of $g$, which
make the surface term vanish when integrating eq.(\ref{E:OV1OV2})
by parts, and of the special polynomial dependence of
$D_{V_1,V_2}(\xi)$ on $\xi$.

\subsection{The geometric term.}
\label{SS:geometricterm}
As we have shown schematically, the
area-scaling properties of correlations depend on the properties
of the geometric term defined in (eq.(\ref{E:defofDV1V2})). In
this subsection we shall study it in some detail.

First we note that there exists a $\xi_{min}$ and $\xi_{max}$,
such that $D(\xi)=0$ for $\xi \geq \xi_{max}$ or $\xi \leq
\xi_{min}$: define
    $\xi_{min} =
        \inf \{
            |\vec{x}-\vec{y}|
            \big|
            \vec{x} \in V_1, \vec{y} \in V_2
            \}$.
For $\xi<\xi_{min}$ there are no values of $\vec{x}$ and $\vec{y}$
which will have a non zero contribution to the integral, and
therefore $D(\xi)=0$ for this region. Similarly
    $\xi_{max} =
        \sup \{
            |\vec{x}-\vec{y}|
            \big|
            \vec{x} \in V_1, \vec{y} \in V_2
            \}$.

For the cases where $\xi_{min}=0$ (that is, the closed sets $V_1$
and $V_2$ are not disjoint), we shall show that $D_{V_1,V_2}(\xi)$
satisfies
\[
    D_{V_1,V_2}(\xi)=G_V V \xi^{d-1}+
        G_S\left( S(B_{in}) - S(B_{out})\right)\xi^d
        +\mathcal{O}(\xi^{d+1}).
\]
$G_V$ and $G_S$ are constants which depend only on the
dimensionality of space, explicitly $G_S = \frac{d\
\pi^{d/2}}{(d-1)\Gamma(d/2+1)}$. $B_{in/out}$ is the common
boundary of $V_1$ and $V_2$. $B_{out}$ is the part of the boundary
when $V_1$ and $V_2$ are on opposite sides of the boundary,
whereas $B_{in}$ is the part of the boundary when $V_1$ and $V_2$
are on the same side of the boundary. The first term is a volume
dependent term.

\begin{figure}[btp]
  \begin{center}
  \scalebox{0.5}{\includegraphics{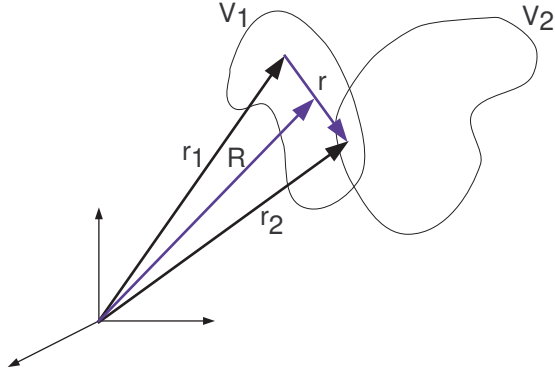}}
  \caption{A pictorial description of the
  ($\vec{r}_1$,$\vec{r}_2$)
  coordinates, and the ($\vec{R}$,$\vec{r}$) coordinates.}
  \label{F:def_of_coords}
  \end{center}
\end{figure}

To solve the integral (\ref{E:defofDV1V2}) we switch to a `center
of mass' coordinate system (shown in Fig.~\ref{F:def_of_coords}):
$\vec{R}=\frac{\vec{r}_1+\vec{r}_2}{2}$, and
$\vec{r}=\vec{r}_2-\vec{r}_1$, so that $\vec{r}$ points from
$\vec{r}_1$ to $\vec{r}_2$, and $\vec{R}$ points to the middle of
$\vec{r}$ (since $\vec{R}=\vec{r}_1+1/2 \, \vec{r}$). In this new
coordinate system,
\begin{equation}
\label{E:DV1V2integral}
    D_{V_1,V_2}(\xi)=
        \int \int
        \delta(\xi-r)
        d^dR d^dr.
\end{equation}

In order to carry out the integration, we wish to use a
generalized `radial' coordinate $\rho$ such that $\rho=\rho_0$
will define the boundary $B(V_1)$ of $V_1$, and generalized
`angular' coordinates $\alpha_i$ which will define solid angles on
the boundary. To define such a coordinate system we foliate space
into surfaces which, when very close to $B(V_1)$, are parallel to
it. $\rho$ is chosen to be the coordinate which points to
different leaves of the foliation. $\rho=\rho_0$, as stated,
defines the surface $B(V_1)$. We also choose $\rho$ such that for
a point $\vec{R}$ for which $|\rho-\rho_0|$ is small enough, then
$|\rho-\rho_0|$ will specify the distance of $\vec{R}$ from the
boundary. $\alpha_i$ are generalized angles on each hyper-surface.
The unit volume in this coordinate system is
\[
    d^dR=J(\rho,\alpha_i)d\rho \prod_i d\alpha_i.
\]

For the vector $\vec{r}$ we choose a polar coordinate system:
\[
    d^dr=r^{d-1}d\Omega=
    r^{d-1} \sin^{d-2}\theta \, d\theta d\Omega_{\bot}.
\]
For a given point $\vec{R}$, the integration over the angular
coordinates of $\vec{r}$ will give us the solid angle subtended by
all allowed values of $\vec{r}_2-\vec{r}_1$, when it is centered
at $\vec{R}$. In preparation for performing the integral in
eq.(\ref{E:DV1V2integral}) for a general geometry (as shown in
Fig.~\ref{F:general case}) we first consider several particular
cases.
\\

{\bf Case 1}: Regions which have a common boundary, but no common
interior. Here we expect to have
\[
    D_{V_1,V_2}(\xi)=
        G_S
        S\left(B(V_1)\cap B(V_2)\right)
        \xi^d
        +\mathcal{O}(\xi^{d+1}).
\]

\begin{figure}[btp]
  \begin{center}
  \scalebox{0.5}{\includegraphics{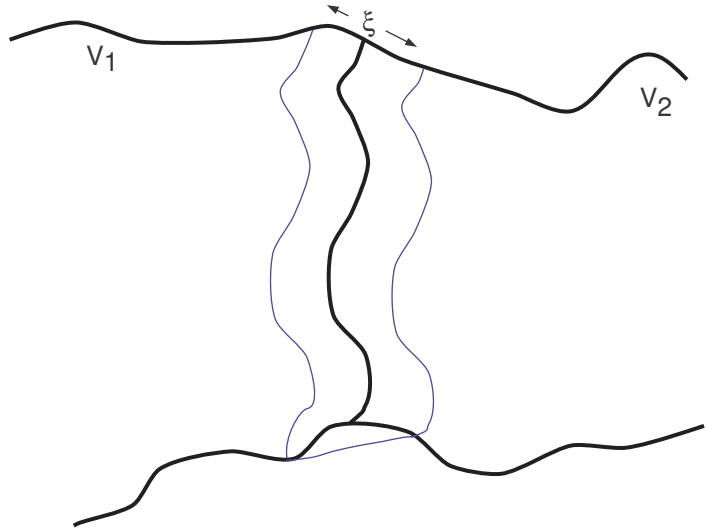}}
  \caption{Region of integration of the $\vec{R}$ coordinate:
  The boundaries of regions 1 and 2 are given by the thick lines.
  The allowed region
  of $\vec{R}$ is given by the thin lines.}
  \label{F:range_of_R}
  \end{center}
\end{figure}

Fixing $\xi$ to have some small value, we look at all values
$\vec{r}$ is allowed to have for a fixed  value of $\vec{R}$.
Defining $\xi=|\vec{r_1}-\vec{r_2}|$, then since $\vec{r}_1$ and
$\vec{r}_2$ are on different sides of the boundary $B\equiv
B(V_1)\cap B(V_2)$, $\vec{R}$ is restricted to a distance of
$\xi/2$ from $B$ (See Fig.\ref{F:range_of_R}). Therefore
\begin{equation}
\label{E:ddRexact}
    \int d^dR =
        \int_{\rho_0-\xi/2}^{\rho_0+\xi/2}
        \int_{A(\rho)}
            J(\rho,\alpha_i)
        d\rho \prod_i d\alpha_i.
\end{equation}
In this case $A(\rho)$ is the region of angular integration for
each leaf of the foliation.

We denote the range of angles which define the common boundary as
$A_B$, so that
\[
    B=\left\{(\rho_0,\alpha_i) | \alpha_i \in A_B\right\}.
\]
We note that with this
definition we may also write
\begin{equation*}
    \int_{A(\rho_0)} J(\rho_0,\alpha_i)\prod_i d\alpha_i=S(B)+ \mathcal{O}(\xi).
\end{equation*}

\begin{figure}[btp]
  \begin{center}
  \scalebox{0.5}{\includegraphics{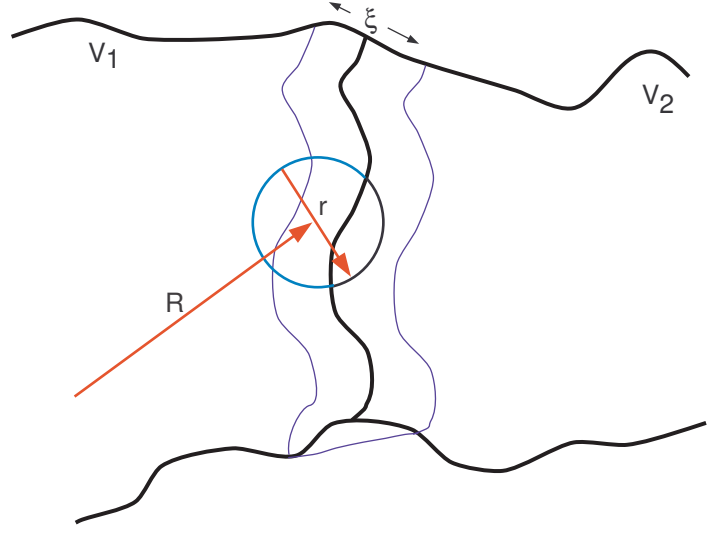}}
  \caption{The allowed region for $\vec{r}$ in the case where the volumes are `just
            touching'. The dark end of the circle shows the region where $\vec{r}$ is allowed
            to point to.}
  \label{F:Allowed_r_1}
  \end{center}
\end{figure}

For a certain point $\vec{R}=(\rho,\alpha_i)$ close to the
boundary: $|\rho-\rho_0|<\xi/2$ (shown in
Fig.~\ref{F:Allowed_r_1}), the angular integration over $\vec{r}$
will be restricted to a region $\bar{\Omega}(\rho,\alpha_i;\xi)$.

Therefore we have:
\begin{equation}
    D_{V_1,V_2}(\xi)=
        \int_{\rho_0-\xi/2}^{\rho_0+\xi/2}
        \int_{A(\rho)}
            J(\rho,\alpha_i)
        \int_{\bar{\Omega}(\rho,\alpha_i;\xi)}
    \xi^{d-1} d\Omega
    \prod_i d\alpha_i
    d\rho.
\end{equation}

Changing variables to the dimensionless variable
$\zeta=(\rho-\rho_0)/(\xi/2)$, and integrating over the angular
coordinate $\Omega$ we obtain,
\begin{equation}
\label{E:dlem1exact}
    D_{V_1,V_2}(\xi)=
        \xi^{d-1}
        \int_{-1}^{1}
        \int_{A(\rho_0+\xi/2\zeta)}
            J(\rho_0+\xi/2\zeta,\alpha_i)
        \bar{\Omega}(\rho_0+\xi/2\zeta,\alpha_i;\xi)
    \prod_i d\alpha_i
    \frac{\xi}{2} d\zeta.
\end{equation}
This is an exact expression.

Expanding the integrand on the r.h.s. of eq.(\ref{E:dlem1exact})
in powers of $\xi$, we obtain
\begin{align}
\notag
        &\int_{A(\rho_0+\xi/2\zeta)}
            J(\rho_0+\xi/2\zeta,\alpha_i)
            \bar{\Omega}(\rho_0+\xi/2\zeta,\alpha_i;\xi)
        \prod_i d\alpha_i\\
\notag
        =&
        \int_{A(\rho_0+\xi/2\zeta)}
            J(\rho_0+\xi/2\zeta,\alpha_i)
            \bar{\Omega}(\rho_0+\xi/2\zeta,\alpha_i;\xi)
        \prod_i d\alpha_i \Big|_{\xi=0}
        + \mathcal{O}(\xi)\\
\label{E:dlem1expansion}
        =&
        \int_{A(\rho_0)}
            J(\rho_0,\alpha_i)
            \Omega(\zeta)
        \prod_i d\alpha_i
        + \mathcal{O}(\xi),
\end{align}
where
\[
\Omega(\zeta)=
    \lim_{\xi\to 0}
    \bar{\Omega}(\rho_0+\xi/2\zeta,\alpha_i;\xi).
\]
We will show that $\Omega$ is a function of $\zeta$ only, and
obtain an explicit expression for it.

\begin{figure}[btp]
  \begin{center}
  \scalebox{0.5}{\includegraphics{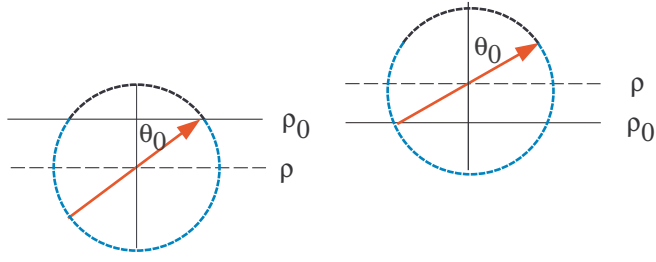}}
  \caption{The allowed region for $\vec{r}$ (dark region) in the case where $\xi$ is very small, so that
    the boundary may be approximated as flat. $\rho_0$ specifies the location of
    the common boundary, and $\rho$ the leaf of the foliation to which $\vec{R}$ is pointing to.
    The left diagram is for the case where $\vec{R} \in V_1$,
    and the right diagram for $\vec{R} \in V_2$.}
  \label{F:Allowed_omega_1}
  \end{center}
\end{figure}

For very small $\xi$, we look at a point $\vec{R}$, a distance
$|\rho-\rho_0|<\xi/2$ from the boundary, and calculate the solid
angle subtended by a vector $\vec{r}$ centered at $\vec{R}$ whose
one end is in $V_1$, and other end is in $V_2$. Since we are
taking the limit where $\xi \to 0$, the shape of the boundary
close to the point $\vec{R}$ may be considered flat (shown in
Fig.~\ref{F:Allowed_omega_1}). Therefore $\Omega$ is the solid
angle generated by $\vec{r}$ when very close to a flat surface.

Defining the z-axis as the axis perpendicular to the surface, we
get that the angle $\theta$ between the z-axis and $\vec{r}$ can
range over values from $0$ to $\theta_0$, with $\cos
(\theta_0)=(\rho_0-\rho)/(\xi/2)$ for $\rho \leq \rho_0$ and $\cos
(\theta_0)=(\rho-\rho_0)/(\xi/2)$ for $\rho \geq \rho_0$. Note
that in any case $0 \leq \theta_0 \leq \pi/2$, implying that $\cos
\theta_0 \geq 0$.

Therefore, for $d>2$,
\[
    \Omega=
    \int_0^{\theta_0} \sin^{d-2}\theta d\theta \int d\Omega_{\bot}.
\]
We note that
\begin{equation}
\label{E:sininttobeta}
    \int_0^{\theta_0} \sin^{d-2}\theta d\theta
    =\frac{1}{2}
        \left(
            B\left(\frac{1}{2},\frac{d-1}{2}\right)-
            B_{\zeta^2}\left(\frac{1}{2},\frac{d-1}{2}\right)
        \right)
\end{equation}
where $B_x(a,b)$ is the partial beta function. The full
solid angle of a d-dimensional sphere is given by
\begin{align}
 \label{stam1}
    C_d&=\int_0^{\pi} \sin^{d-2}\theta d\theta \int d\Omega_{\bot} \notag\\
        &=B\left(\frac{1}{2},\frac{d-1}{2}\right) \int d\Omega_{\bot}.
\end{align}
Putting eq.(\ref{stam1}) into eq.(\ref{E:dlem1expansion}), we get
\[
        \int_{A(\rho_0)}
            J(\rho_0,\alpha_i)
            \frac{C_d}{2}
                \left(
            1-\frac{B_{\zeta^2}(\frac{1}{2},\frac{d-1}{2})}{B(\frac{1}{2},\frac{d-1}{2})}
                \right)
        \prod_i d\alpha_i
        + \mathcal{O}(\xi).
\]
Integrating over the angular coordinates we are left with
\[
    \frac{C_d}{2}
    \left(
        1-\frac{B_{\zeta^2}(\frac{1}{2},\frac{d-1}{2})}{B(\frac{1}{2},\frac{d-1}{2})}
    \right)
    S(B)
    +\mathcal{O}(\xi),
\]
and plugging this into eq.(\ref{E:dlem1exact}), will give us:
\begin{align}
\notag
    D_{V_1,V_2}(\xi)&=\frac{\xi}{2}\xi^{d-1}
        \left[\frac{C_d}{2}
            \int_{-1}^1
                \left(
                    1-\frac{B_{\zeta^2}(\frac{1}{2},\frac{d-1}{2})}{B(\frac{1}{2},\frac{d-1}{2})}
                \right)
                S(B) d\zeta
                +\mathcal{O}(\xi)
        \right]\\
\label{E:lem1complem2}
    &=\frac{\xi^{d}}{2}C_d S(B)
        \int_{0}^1
        \left(
            1-\frac{B_{\zeta^2}(\frac{1}{2},\frac{d-1}{2})}{B(\frac{1}{2},\frac{d-1}{2})}
        \right)
        d\zeta
        +\mathcal{O}(\xi^{d+1}).
\end{align}
So now we have to evaluate the integral
\[
    \int_0^1
        B_{\zeta^2}\left(\frac{1}{2},\frac{d-1}{2}\right)
    d\zeta=
    \int_0^1
        \int_0^{\zeta^2} t^{-\frac{1}{2}}(1-t)^{\frac{d-3}{2}} dt
    d\zeta,
\]
which can be done by changing the order of integration. The result
is that
\begin{equation*}
    D_{V_1,V_2}(\xi)
      =\xi^{d} \frac{C_d}{2} S(B)
        \frac{B(1,\frac{d-1}{2})}{B(\frac{1}{2},\frac{d-1}{2})}
         +\mathcal{O}(\xi^{d+1}).
\end{equation*}

Expressing the Beta function as a product of Gamma functions, and
using the explicit expression
$C_d=\frac{d\pi^{d/2}}{\Gamma(d/2+1)}$, we get the final result
\begin{equation}
 D_{V_1,V_2}(\xi)= \frac{d\pi^{d/2}}{(d-1) \Gamma(d/2+1)} S(B) \xi^d + \mathcal{O}(\xi).
\end{equation}
\\

{\bf Case 2}: Next we consider geometries where $V_1=V_2 \equiv
V$, for which we expect to have:
\[
    D_{V_1,V_2}(\xi)=G_V V \xi^{d-1} - G_S S(B(V)) \xi^d + \mathcal{O}(\xi^{d+1}).
\]

We consider the same foliation of space as before. In this case,
even at small $\xi$, we integrate over points inside $V$. The
angular integration over $\vec{r}$ will be constrained only when
the region where $\vec{R}$ is a distance of $\xi/2$ from the
boundary. Therefore:
\begin{align}
\label{E:lem2dvexact}
    D_{V_1,V_2}(\xi)&=
        \int_0^{\rho_0}
        \int_{A(\rho)}
        \int_{\bar{\Omega}}
            \xi^{d-1} J(\rho,\alpha_i)
        d\Omega \prod_i
        d\alpha_i
        d\rho\\
\notag
    &=\int_0^{\rho_0-\xi/2}
        \int_{A(\rho)}
            \xi^{d-1} J(\rho,\alpha_i)
        C_d
        \prod_i d\alpha_i
        d\rho\\
\notag
    &+\int_{\rho_0-\xi/2}^{\rho_0}
        \int_{A(\rho)}
        \int_{\bar{\Omega}}
            \xi^{d-1} J(\rho,\alpha_i)
        d\Omega \prod_i
        d\alpha_i
        d\rho.
\end{align}
The first expression is the unconstrained part, and the second
expression is the constrained part.

Considering first the integral for which $\vec{r}$ is not
constrained, we have
\begin{multline*}
    \int_0^{\rho_0-\xi/2}
        \int_{A(\rho)}
            \xi^{d-1} J(\rho,\alpha_i)
            C_d
        \prod_i d\alpha_i
        d\rho\\
    =V C_d \xi^{d-1}
        - \int_{\rho_0-\xi/2}^{\rho_0}
            \int_{A(\rho)}
            \xi^{d-1} J(\rho,\alpha_i)
            \int_0^{\pi}\sin^{d-2}\theta d\theta d\Omega_{\bot}
        \prod_i d\alpha_i
        d\rho.
\end{multline*}
Proceeding as before, we change variables of integration to
$\zeta=(\rho-\rho_0)/(\xi/2)$, and expand in small $\xi$.
\begin{equation}
\label{E:lem2vcont}
        V C_d \xi^{d-1}
        - S(B) \frac{\xi^{d}}{2}
        \int_{-1}^{0}
            \int_0^{\pi}\sin^{d-2}\theta d\theta d\Omega_{\bot}
        d\zeta
        +\mathcal{O}(\xi^{d+1}).
\end{equation}

Under the same approximation, the constrained part reduces to
\begin{equation}
\label{E:lem2scont}
    S(B) \frac{\xi^d}{2}
    \int_{-1}^{0}
        \Omega(\zeta)
    d\zeta
    +\mathcal{O}(\xi^{d+1}).
\end{equation}

In this case, $\theta$, the angle between $\vec{r}$ and the
z-axis, is restricted to $\theta_0$, and $\pi-\theta_0$ where
$\cos\theta_0=-\zeta$. Hence
\[
    \Omega(\zeta)=
        \int_{\theta_0}^{\pi-\theta_0}
            \sin^{d-2}\theta
        d\theta.
\]

Combining the contributions of the surface terms
of (\ref{E:lem2vcont}) and (\ref{E:lem2scont}) to
those of the surface term of (\ref{E:lem2dvexact}), we have
\[
    - S(B) \frac{\xi^d}{2}
    \int_{-1}^{0}
        \left(
        \int_0^{\theta}+
        \int_{\pi-\theta}^{\pi}
        \right)
            \sin^{d-2}\theta d\theta
        \int \Omega_{\bot}
     d\zeta.
\]
Since $0 \leq \theta \leq \pi/2$, both angular integrals are
equal, and the above equation simplifies to
\[
    - S(B) \frac{\xi^d}{2}
    \int_{-1}^{0}
        2
        \int_0^{\theta}
            \sin^{d-2}\theta d\theta
        \int \Omega_{\bot}
    d\zeta.
\]
Using eq.(\ref{E:sininttobeta}) we may carry out the integral over
the $\theta$ coordinate:
\begin{multline*}
    - S(B) \frac{\xi^d}{2}
    2\int_{-1}^{0}
        \frac{1}{2}
        \left(
            B\left(\frac{1}{2},\frac{d-1}{2}\right)-
            B_{\zeta^2}\left(\frac{1}{2},\frac{d-1}{2}\right)
        \right)
    d\zeta\\
    =
    - S(B) \frac{\xi^d}{2}
    \int_{-1}^{1}
        \frac{1}{2}
        \left(
            B\left(\frac{1}{2},\frac{d-1}{2}\right)-
            B_{\zeta^2}\left(\frac{1}{2},\frac{d-1}{2}\right)
        \right)
    d\zeta.
\end{multline*}

Comparing this with eq.(\ref{E:lem1complem2}), we find that the
surface term contribution has exactly the same magnitude as the
leading order contribution to $D_{V_1,V_2}(\xi)$ in Case 1, its
sign, however, is negative.
\\

{\bf Case 3}: $V_1$ is fully contained in $V_2$ with no common
boundaries. Here we shall have:
\[
    D_{V_1,V_2}=C_d V_1\xi^{d-1}+\mathcal{O}(\xi^{d+1}).
\]
Consider $V_1$ and its complement in $V_2$: $V_2\setminus V_1$. We
may write $D_{V_1,V_2}(\xi)=D_{V_1,V_1}(\xi)+D_{V_1,V_2\setminus
V_1}(\xi)$. $V_1$ and $V_2\setminus V_1$ are disjoint sets with a
common boundary, therefore they satisfy the conditions of Case 1.
Applying the results of Case 1 to $V_1$ and $V_2\setminus V_1$ we
get
\begin{align*}
    D_{V_1,V_2 \setminus V_1}(\xi)&=
        G_S \xi^d S\left(B(V_1) \cap B(V_2\setminus V_1)\right)+\mathcal{O}(\xi^{d+1})\\
        &=
        G_S \xi^d S\left(B(V_1)\right)+\mathcal{O}(\xi^{d+1}).
\end{align*}
Using Case 2 to calculate $D_{V_1,V_1}(\xi)$, we find that the
surface terms in $D_{V_1,V_1}(\xi)+D_{V_1,V_2\setminus V_1}(\xi)$
exactly cancel each other.
\\

{\bf Case 4}: $V_1$ is contained in $V_2$ and they do have a
common boundary. In this case the result is that
\[
    D_{V_1,V_2}=C_d V_1 \xi^{d-1}-G_S S(B(V_1)\cap B(V_2)) \xi^{d}+\mathcal{O}(\xi^{d+1}).
\]

Defining the common boundary of $V_1$ and $V_2$ as $ B \equiv
B(V_1) \cap B(V_2) $, we consider $V_3$, the complement of $V_2$.
Since $V_2$ and $V_3$ have the same boundary then $B(V_1) \cap
B(V_2) = B(V_1) \cap B(V_3) = B$. Since the interior of $V_3$ is
disjoint from the interior of $V_1$ ($V_1 \subset V_2$), it
satisfies the conditions for Case 1, so that $D_{V_1,V_3}(\xi) =
G_S \, S(B(V_1)\cap B(V_2)) \xi^{d}+\mathcal{O}(\xi^{d+1})$. Using
Case 3, we also have that $D_{V_1,V_2\cup V_3}(\xi)=C_d V_1
+\mathcal{O}(\xi^{d+1})$, so that $D_{V_1,V_2}(\xi)=C_d V_1 -
D_{V_1,V_3}(\xi)+\mathcal{O}(\xi^{d+1})$, which gives the deired
result.
\\

{\bf Case 5}: $V_1$ and $V_2$ have no common boundary and are not
disjoint.
\[
    D_{V_1,V_2}=C_d (V_2\cap V_1) \xi^{d-1}+\mathcal{O}(\xi^{d+1}).
\]

We have already seen that this is correct if $V_1$ is fully
contained in $V_2$. What is left, is to check the case when only
part of $V_1$ is contained in $V_2$. In this case we may define
$V_{2,in}=V_2 \cap V_1$, and $V_{2,out}=V_2 \setminus V_{2,in}$
(so that $V_2=V_{2,in} \cup V_{2,out}$). The boundary of
$V_{2,in}$ has a common boundary with $V_1$. Since $V_{2,out}$ is
the complement of $V_{2,in}$ relative to $V_2$, any boundary of
$V_{2,out}$ which is not a boundary of $V_2$ is also a boundary of
$V_{2,in}$, and since $V_2$ does not have a common boundary with
$V_1$, any boundary of $V_{2,in}$ which is common to $V_1$ must
also be common to $V_{2,out}$. Therefore to leading order,
\begin{align*}
    D_{V_1,V_2}&=D_{V_1,V_{2,in}}+D_{V_1,V_{2,out}} \\
               &= C_d V_{2,in} \xi^{d-1}
                    + \frac{C_{d-1}}{d-1}
                      \left(
                        S(B(V_{2,out}) \cap B(V_1))-S(B(V_{2,in}) \cap B(V_1))
                      \right) \xi^d\\
               &=C_d (V_2 \cap V_1) \xi^{d-1}.
\end{align*}
We are using here somewhat imprecise notation as we are not
differentiating between the set $V_1 \cap V_2$ and its volume.\\

{\bf General geometries}: In order to avoid problems with
infinities we consider:
\begin{enumerate}
\item
    Regions $V_1$ and $V_2$ whose common
    boundary has a finite number of connected components.
\item
    Both volumes are connected \footnote{We state this in order to avoid
    the case where the volumes are composed of infinitely many
    disconnected subsets. If the volumes are composed of a finite number of disconnected
    subsets, then this case can easily be reduced to the connected volumes
    case.}.
\end{enumerate}
An example of regions satisfying these conditions is given in
Fig.~\ref{F:general case}.

The idea is to divide $V_2$ into subsets, such that each subset of
$V_2$ will either contain a single connected subset of the
boundary of $V_1$ which is common to $V_2$, or no such boundary at
all: consider a single connected part of $B(V_1) \cap B(V_2)$,
which we denote by $B$. We shall construct a volume $V_2^{\prime}
\subseteq V_2$ such that (a) $B = B(V_1) \cap B(V_2^{\prime})$ and
(b) $V_2^{\prime}$ has no other common boundaries with $V_1$, and
does not contain a boundary of $V_1$.

This construction is achieved by first enclosing $B$ within a
volume $A$ such that $A$ does not contain, and is far enough from,
any other boundary of $V_1$. Then we define $V_2^{\prime} = A \cap
V_2$. That $V_2^{\prime} \subseteq V_2$ is obvious from the
definition. Condition (a) is satisfied since $B$ is common to both
$B(V_2)$ and $B(A)$. Condition (b) is satisfied since if
$V_2^{\prime}$ has another boundary with $V_1$ or contains a
boundary of $V_1$ then this must be contained in $A$ or a boundary
of $A$, which is a contradiction.

This procedure can be applied consecutively to all common
components of the boundary of $V_1$ and $V_2$, in such a way that
all the $V_2^{\prime}$ are disjoint (if they are not disjoint we
may get rid of their common part by a redefinition). Thus we
obtain a finite collection of sub-volumes $V_{2_i}$ that are
contained in $V_2$, and that  satisfy $\left( \bigcup_i B(V_{2_i})
\right) \cap B(V_1) = B(V_2) \cap B(V_1)$. In words:  the common
boundary of the $V_{2_i}$'s with $V_1$ is equal to the common
boundary of $V_2$ and $V_1$.

We also define $V_{2,bulk}=V_2 \setminus \bigcup_i V_{2_i}$, so
that $V_2 = \bigcup_i V_{2_i} \cup V_{2,bulk}$. $V_{2,bulk}$ will
have no common boundary with $V_1$. This follows from observing
that if it does have a common boundary with $V_1$ then this
boundary must not be a boundary of $V_2$ with $V_1$ since all of
these appear in the $V_{2_i}$'s. Since $V_{2,bulk}$ is the
complement of $\bigcup_i V_{2_i}$ with respect to $V_2$, then
$\bigcup_i V_{2_i}$ must also have a boundary common to $V_1$
which is not common to $V_2$, but this is not allowed by
construction.

Now, $D_{V_1,V_{2, bulk}} \propto V_1\cap V_{2,bulk} +
\mathcal{O}(\xi^{d+1})$ since it satisfies the conditions in Case
5. Also we note that for each $i$, the interior of $V_{2_i}$ is
either contained in $V_1$ or disjoint from $V_1$---an observation
which follows from the fact that no interior of $V_{2_i}$ contains
a boundary of $V_1$ and so cannot cross from the interior to the
exterior of $V_1$.

This implies that for indices $i$ for which $V_{2_i} \subseteq
V_1$, we get
\begin{align*}
    \sum_{i \in in} S(B(V_{2_i})\cap B(V_1))
        &=S\left(
                \bigcup_{i \in in} \left(V_{2_i} \cap B(V_1)\right)
            \right)\\
        &= S\left(
                \left(\bigcup_{i \in in} V_{2_i} \right) \cap B(V_1)
            \right)\\
        &\equiv S_{in}
\end{align*}
Where $S_{in}$ is the surface area of the common boundary of $V_2$
and $V_1$ for which, close to the boundary, the interiors are not
disjoint. $S_{out}$ is similarly defined.

We also note, that using case 4, $D_{V_1,V_{2_i}} \sim V_{2_i} \pm
S(B(V_{2_i})\cap B(V_1))$, for $i \in in$ (or $i \in out$).
Therefore, the leading order behavior of $D_{V_1,V_2}=\sum_i
D_{V_1,V_{2_i}}+D_{V_1,V_{2,bulk}}$ is
\begin{multline*}
    C_d V_{2,bulk} \cap V_1 \xi^{d-1}+
\\
    \frac{C_{d-1}}{d-1}
        \left(
            \sum_{i \in in} (V_{2_i} \cap V_1 + S(B(V_{2_i})\cap B(V_1)))
          - \sum_{j \in out} (S(B(V_{2_j})\cap B(V_1)))
        \right) \xi^d,
\end{multline*}
which reduces to
\[
               C_d V_2 \cap V_1 \xi^{d-1}
                 + \frac{C_{d-1}}{d-1}
                 \left(S_{in} - S_{out}\right)
                 \xi^d.
\]
This completes the proof of our claim regarding the leading order
behavior of the geometric term $D_{V_1,V_2}(\xi)$.

\subsection{Area-scaling of two point functions.}
Going back to eq.(\ref{E:OV1OV2}), we can now evaluate:
\begin{equation*}
    \langle O^{V_1}_i O^{V_2}_j\rangle_C =\int_{\xi_{min}}^{\xi_{max}}
    D(\xi)F(\xi)
    =\int_{\xi_{min}}^{\xi_{max}} D(\xi)
        \nabla^2
        g
        d\xi.
\end{equation*}

First we integrate by parts (See eq.(\ref{E:partshandwaving})),
\begin{align*}
    \langle O^{V_1}_i O^{V_2}_j\rangle_C
      &= \int D(\xi) \frac{1}{\xi^{d-1}} \frac{d}{d \xi} \xi^{d-1} \frac{d}{d \xi} g(\xi) d\xi \\
      &= D(\xi) \frac{d}{d \xi} g(\xi) \big|_{\xi_{min}}^{\xi_{max}}
       - \int \frac{d}{d \xi} \left( D(\xi) \frac{1}{\xi^{d-1}}\right)
         \xi^{d-1} \frac{d}{d \xi} g(\xi) d\xi.
\end{align*}
Consider the surface term. For finite and non zero $\xi_{min}$ and
$\xi_{max}$, $D$ vanishes and therefore the surface term vanishes.
When $\xi_{min}=0$, we note that at small $\xi$, since $D(\xi)
\sim V \xi^{d-1} + \mathcal{O}(\xi^{d})$, then
\[
    D g^{\prime} \sim V \xi^{d-1} \frac{1}{\xi^{a-1}} +\mathcal{O}(\xi^{d-a+1}),
\]
and because of condition (I) this term vanishes. We shall see that
the limit $\xi_{max} \to \infty$ poses a special problem, and may
require the introduction of a long-distance (IR) cutoff.

We now have:
\[
    \langle O^{V_1}_i O^{V_2}_j\rangle_C =
        - \int_{\xi_{min}}^{\xi_{max}} \widetilde{D}(\xi) \xi^{d-1} g^{\prime}(\xi) d\xi,
\]
where we have defined
$\widetilde{D}(\xi) = \frac{d}{d \xi} \left(D(\xi)
\frac{1}{\xi^{d-1}}\right)$.
Since $\widetilde{D}$ is constant for small $\xi$ then in order
for $\langle O^{V_1}_i O^{V_2}_j\rangle_C$ to converge at the
lower limit, we need that condition (II) be satisfied.

Introducing a short-distance (UV) cutoff scale $\Lambda$,
\begin{align}
\label{stam2}
    \langle O^{V_1}_i O^{V_2}_j\rangle_C &=
        - \int_{\xi_{min}}^{\xi_{max}}
            \Lambda^{-(d-1)} \widehat{D}(\Lambda \xi) \Lambda^{-(d-1)} (\Lambda\xi)^{d-1}
            \Lambda^{\alpha} g^{\prime}(\Lambda \xi) \Lambda^{-1}
          d\Lambda \xi \notag \\
      & = - \int_{\xi_{min}}^{y_{max}} \Lambda^{-2d+1+\alpha} \widehat{D}(y) y^{d-1} g^{\prime}(y)
      dy,
\end{align}
where $\alpha$ is the scaling dimension of $g^{\prime}$:
$\Lambda^{\alpha} g^{\prime}(\Lambda \xi;\Lambda^t m,1) =
    g^{\prime}(\xi;m,\Lambda).$
Here we have introduced explicitly the parameter `m' to allow for
the possibility that the theory contains other dimensionful
parameters in addition to $\Lambda$ (such as masses, or an IR
cutoff). For simplicity, we have introduced a single such
parameter. $\widehat D$ is dependent on the (dimensionful)
geometric parameters of $V_1$ and $V_2$. An explicit expression
for $\alpha$ is obtained by noting that if
$\mathcal{O}_i(\vec{x})$ has scaling dimension $\delta_i$, then
$F$ has dimension $\delta_i+\delta_j+2d$, so that $g^{\prime}$ has
dimension $\alpha = \delta_i+\delta_j+2d-1 \equiv \delta+2d-1$.
Also, $\widetilde{D}$ scales as $[\text{length}]^{d-1}$, so that
the geometric parameters in $\widehat{D}$ are now scaled to a
length $1/\Lambda$.

Taking the shortest scale in the problem to be $1/\Lambda$, we
wish to take the limit $\Lambda \xi \to \infty$. When both regions
are disjoint and $\xi_{min} \neq 0$, we find that the integral in
eq.(\ref{stam2}) vanishes. This is supported by numerical results
where it is seen that disjoint volumes have zero covariance as
long as the distance between them is larger than the UV scale
\cite{Implications}. If the regions have some overlap or a common
boundary then $\xi_{min} = 0$. Since $\widetilde{D}(\xi)$ is
constant at small $\xi$ then $\widehat{D}(y)$ will be constant for
all but very large $y$. This will then allow us to evaluate
$\langle O^{V_1}_i O^{V_2}_j\rangle_C$ as
\begin{equation*}
    \langle O^{V_1}_i O^{V_2}_j\rangle_C = \Lambda^{\delta}
            \int_0^{\infty} \widehat{D}(y) y^{d-1} g^{\prime}(y) dy +\mathcal{O}(1/y_{max})
\end{equation*}
as long as there are no contributions from the `rescaled
infinity'. Writing out $\widetilde{D}(\xi)=\sum_{n=0} d_n \xi^n$,
or $\widehat{D}(y)=\sum_{n=0} d_n/\Lambda^{-(d-1)+n} y^n$, we get
that
\begin{equation}
\label{E:finalresult}
    \langle O^{V_1}_i O^{V_2}_j\rangle_C =
        - \Lambda^{\delta} \Lambda^{d-1} d_0 \int_0^{\infty} y^{d-1} g^{\prime} (y) dy
        +\mathcal{O}(1/\Lambda \xi_{max}, d_n/\Lambda^{-(d-1)+n}, \Lambda^t m).
\end{equation}
$d_0$ is the mutual surface area. We have established that
$\langle O^{V_1}_i O^{V_2}_j\rangle_C$ scales linearly with the
area of the common boundary, to leading order in the geometric
parameters. The remaining terms in eq.(\ref{E:finalresult}) that
contain additional  geometric parameters, such as $\xi_{max}$, and
$d_n$ (for $n>0$), are subleading and scale with a smaller power
of the area. For example, if the spatial regions are
$d$-dimensional spheres of radii $R$ then the subleading terms
scale as $R^a$ with $a<d-1$.

We would like to point out that the leading power of $\Lambda$ in
the expansion of $\langle O_i^V O_j^V\rangle_C$ may appear in one
of the subleading terms, where it would appear multiplied by one
of the other dimensionful parameters which we have denoted by $m$.
Indeed, introducing an IR scale $L$ in the theory may spoil the
area-scaling behavior. We shall show an example of this in the
next section.

\section{Explicit calculations.}
\label{S:Explicit_Cal}
We have carried out several independent calculations of the energy
fluctuations, and `boost generator fluctuations' of a free massive
scalar field in various volumes
\cite{Shortest,Implications,Dprivate,Sprivate,TandA}. We present
some of these calculations here.

\subsection{The two point function of the energy operator}
As an
explicit example, we wish to find the two point function for the
energy operator in a volume $V$ of Minkowski space for a free
massless scalar field. This is given by
\begin{multline}
\label{E:HV2}
    <:E^{V_1}::E^{V_2}:>=\frac{1}{8} \frac{1}{(2\pi)^{2d}}
              \int \int \int_{V_1} \int_{V_2}
                {\left(\frac{-\vec{p} \cdot \vec{q}}
                    {\sqrt{\omega_p \omega_q}} -
                    \sqrt{\omega_p \omega_q}\right)}^2\\
                \times e^{\imath(\vec{p}+\vec{q})\cdot (\vec{x}-\vec{y})}
                d^dp\,d^dq\,d^dx\,d^dy,
\end{multline}
which may be written in the form $
    \int \limits_0^\infty F(\xi) D(\xi) d\xi,
$ where for a free field theory,
\begin{equation}
 \label{fx19}
    F(x)=\frac{1}{8} \frac{1}{(2\pi)^{2d}}
           \int
            \left( pq+
                2\vec{p} \cdot \vec{q}+
                \frac{(\vec{p} \cdot {\vec{q}})^2}{pq} \right)
            e^{-\imath(\vec{p}+\vec{q})\cdot \vec{x}}
                d^dp\,d^dq.
\end{equation}

To find an expression for $F(\xi)$, we switch to a coordinate
system where:
\[
    \vec{x}=
        \begin{pmatrix}
        x \\
        0 \\
        0 \\
        \vdots \\
        0 \\
        \end{pmatrix}
    ;\,
    \vec{q}=
        \begin{pmatrix}
        q_x \\
        q_{\bot} \\
        0 \\
        \vdots \\
        0 \\
        \end{pmatrix}
        ;\,
    \vec{p}=
        \begin{pmatrix}
        p_x \\
        p_{\bot} \cos \theta_p \\
        p_{\bot_{\bot}} \\
        \vdots \\
        0 \\
        \end{pmatrix}.
\]
In this form, we may do all angular integrations:
\begin{multline*}
    F(x)=\frac{1}{8} \frac{1}{(2\pi)^{2d}}
           \left(
                \frac{\pi^{\frac{d}{2}}}{\Gamma(\frac{d}{2}+1)}(d-1)
           \right)^2\\
           \times
           \int
                \left( pq+
                    2p_xq_x+
                    \frac{p_x^2q_x^2}{pq}+
                    \frac{p_{\bot}^2q_{\bot}^2}{pq}\frac{1}{d-1}
                \right)\\
            \times
                e^{-\imath(p_x+q_x)x}
                p_{\bot}^{d-2}
                q_{\bot}^{d-2}
            dp_{\bot}\,dq_{\bot}
            dp_x\,dq_x.
\end{multline*}
Switching to polar coordinates in the remaining two dimensional
system:
\begin{align*}
    p_{\bot}&=p \sin \theta \\
    p_x&=p \cos \theta,
\end{align*}
and noting that integrations over the $p$ and $q$ variables are
independent, we can now evaluate the integral by imposing an
exponential cutoff $C(p/\Lambda)=e^{-p/\Lambda}$.
\begin{align}
 \label{fx20}
    F(x)&=\frac{(d+1)
                  \Gamma\left(\frac{d+1}{2}\right)
                  \Lambda^{2(d+1)}}
                 {8 \pi^{d+1}
                  (1+(\Lambda x)^2)^{d+3}}
            (d-2(d+2)(\Lambda x)+d(\Lambda x)^4 \notag \\
          &=\frac{(d+1)
                  \Gamma\left(\frac{d+1}{2}\right)
                  \Lambda^{2(d+1)}}
                 {8 \pi^{d+1}}
            \nabla^2_{\Lambda x}
            \frac{(\Lambda x)^2-1)}
                 {2(d+2)(1+(\Lambda x)^2)^{d+1}}.
\end{align}

Therefore, for volumes with a common boundary B,
\[
    \langle :E^{V_1}::E^{V_2}:\rangle
    \approx
    -\frac{(d+1) \Gamma^2\left(\frac{d+1}{2}\right) \Lambda^{d+1}}
          {8\pi^{d+1}}
    \frac{C_{d-1}}{d-1}
    \int_0^{\infty}
        y^{d-1}
        \frac{\partial}{\partial y}
        \frac{y^2-1}{2(d+2)(1+y^2)^{d+1}}
    dy.
\]
Doing this integral we get:
\[
    \langle :E^{V_1}::E^{V_2}:\rangle
    \approx
    -\frac{\Gamma\left(\frac{d+1}{2}\right)
           \Gamma\left(\frac{d+3}{2}\right)}
          {\Gamma\left(2+\frac{d}{2}\right)}
    \frac{\Lambda^{d+2}}{2^{d+4}\pi^{\frac{d}{2}+1}}
    (S(B_{out})-S(B_{in})).
\]

\subsection{The Boost operator for half of Minkowski space.}
As another exercise, we calculate the variance of two boost
operators, when the volume in question is half of Minkowski space.
We start with fluctuations of the boost generator in the `z'
direction, ${B^{(z)}}^V$, where \cite{weinbergfieldtheory1}
\[
    {B^{(z)}}^V=\int \limits_{V} z\mathcal{H}d^dx.
\]

Now,
\[
    <:({B^{(z)}}^V)^2:>=\int z_1 z_2 F_{E,E}(|\vec{x}_1-\vec{x}_2) d^dx_1\,d^dx_2.
\]
In this case it will be more useful to calculate $D(\xi)$
explicitly:
\[
    D_d(\xi)=\int \limits_{-\infty}^{\infty}
             \ldots
             \int \limits_{-\infty}^{\infty}
             \int \limits_0^\infty
             \int \limits_0^\infty
             \delta^{(d)}(\xi-|\vec{x}_1-\vec{x}_2|)
             d^dx_1 \, d^dx_2.
\]
Switching to $\vec{r}_{\pm}=\vec{x}_1 \pm \vec{x}_2$ coordinates,
we may integrate over the transverse $\vec{r}_+$ directions,
yielding the transverse volume $V_{\bot}$ (Transverse meaning
the direction transverse to the $z$ coordinate.) Therefore:
\[
    D_d(\xi)=\frac{1}{2} V_{\bot}
             \int \limits_0^\infty
             \int \limits_{-z_+}^{z_+}
             \left[
                \int \limits_{-\infty}^{\infty}
                \ldots
                \int \limits_{-\infty}^{\infty}
                \delta^{(d)}(\xi-r_-)
                d^{d-1}{r_-}_{\bot}
             \right]
             dz_-
             dz_+.
\]
Hence
\begin{equation}
\label{E:Boostflucs1}
    <:({B^{(z)}}^V)^2:>
        =V_{\bot}
         \int \limits_0^{\infty}
         \int \limits_{-z_+}^{z_+}
         \left[
            \int \limits_{-\infty}^{\infty}
            \ldots
            \int \limits_{-\infty}^{\infty}
            \frac{1}{4}(z_+^2-z_-^2)
            F_d(r_-)
            d^{d-1}{r_-}_{\bot}
         \right]
    dz_-\,dz_+,
\end{equation}
where $F(r_-)$ is defined in eq.(\ref{fx19}) and is evaluated in
eq.(\ref{fx20}).

By integrating by parts, this integral may be carried out exactly
\[
    <:({B^{(z)}}^V)^2:>=\frac{1}{\Lambda^2}
                      \frac{1}{d-1}
                      <:E_V:^2>.
\]
Here, again, the fluctuations are proportional to the surface
area.

When we consider boosts in the directions parallel to the
boundary, we need to deal with the IR scale `$L$' (which needs to
be introduced in order to define the boundary). Treating the IR
scale as a dimensional parameter of dimension -1, we see that
according to (\ref{E:finalresult}) it may contribute a factor of
$L$ to the fluctuations. It is a simple matter to generalize
equation (\ref{E:Boostflucs1}) to boosts in the other directions,
yielding
\[
    <:({B^{(\bot)}}^V)^2:>=V_{\bot} L^2 \Lambda^{d+1}
                         \frac{(d+1) \Gamma^2 \left( \frac{d+1}{2} \right)}
                              {16^{d+6} \pi^{\frac{d}{2}+1}
                                \Gamma \left(2+\frac{d}{2}\right)}
                         +\mathcal{O}(V_{\bot}).
\]
We get
that the fluctuations are not proportional to the surface area. As
stated earlier this is due to the IR cutoff we imposed on
directions parallel to the boundary of half of space.

\section{Discussion}
\label{S:Discussion}
We have shown that under conditions (I) and (II), two point
functions  of bulk operators restricted to some regions of
Minkowski space will scale as the surface area of the common
boundary of the two regions, independently of their geometries.

Generally, one would expect that fluctuations, quantum or
statistical, scale as the volume in which they are being measured,
and not as the surface area of the volume. In a thermodynamic
context, fluctuations of observables represent thermodynamic
quantities. Energy fluctuations for example, correspond to heat
capacity, which is usually extensive.

We have found instead, that fluctuations scale as the area of the
region of space in which they are being measured. We can give a
thermodynamic interpretation to the quantum fluctuations that we
have calculated by considering not an observer making quantum
measurements inside the volume $V$, but a different observer who
has no access to the region outside $V$. If the initial state of
the system in the whole space is $| \psi \rangle$, then an
observer that has no access to the region outside $V$ will see a
state described by the density matrix
    $\rho_{in}=\text{Trace}_{out} |\psi \rangle \langle \psi |$.
It is possible to show \cite{Feynmanstatmech} that for any
operator $O^{V}$ which acts only inside the region $V$,
    $\langle \psi | O^{V} | \psi \rangle = \text{Trace}(\rho_{in}O^{V})$.
Therefore, the quantum fluctuations seen by the first observer
(which have area-scaling properties), are the same as the
statistical fluctuations seen by the second observer. For the
latter observer, fluctuations of, say, the energy, are a measure.
of the heat capacity, which according to the above result is
proportional to the surface area of $V$ and not to its volume. The
fluctuation-dissipation theorem then generalizes this result for
fluctuations of other operators, implying that thermodynamic
quantities have area-scaling behavior \cite{TandA}.

Going one step further, if the volume $V$ is chosen to be half of
Minkowski space, then $\rho_{in}$ is the density matrix for an
accelerated observer \cite{holzheyetal, KabStr}. One can calculate
the heat capacity of Rindler space radiation by the above method,
yielding, again, an area dependent quantity. This is consistent
with area-scaling behavior of Unruh radiation \cite{TandA}.

These area dependent fluctuations also give evidence for a
boundary type theory. If one considers correlations between two
operators $\langle O^{V_1}_i O^{V_2}_j \rangle$, then these
correlations vanish when no common boundary exists---Implying that
the information content of the system exists on the boundary. This
line of thought is further developed in \cite{Implications}.

It is interesting to note that since, in general, $\langle
0|O^{V_1} O^{V_2} |0\rangle \neq \langle0| O^{V_1}|0\rangle
\langle 0| O^{V_2} |0 \rangle$, then when $V_2$ is the complement
of $V_1$, we get that the vacuum is an entangled state (see
\cite{Reznik} for a discussion of this). This implies, due to Bell
inequalities, that such correlation functions cannot be reproduced
by a local classical setup.

Finally, a proposed application of this area-scaling behavior was
recently given by \cite{OakninHZ} to explain the
Harrison-Zeldovich spectrum.

\section{acknowledgements}
Research supported in part by the Israel Science Foundation under
grant no. 174/00-2 and by the NSF under grant no.PHY-99-07949.
A.~Y. is partially supported by the Kreitman foundation.  R.~B.
thanks the KITP, UC at Santa Barbara, where this work has been
completed. We would like to thank  M. Einhorn, B. Kol, and M.
Srednicki for useful discussions, and D. Oaknin for many
interesting, detailed and helpful discussions on all aspects of
this paper.

\end{document}